\documentstyle[11pt,newpasp,twoside,epsfig]{article}
\def\cm2{cm$^{-2}$}
\newcommand{\lsim}{\ \raise -2.truept\hbox{\rlap{\hbox{$\sim$}}\raise5.truept
        \hbox{$<$}\ }}
\newcommand{\gsim}{\ \raise -2.truept\hbox{\rlap{\hbox{$\sim$}}\raise5.truept
        \hbox{$>$}\ }}
\def\nmea{69}  % DLAs with measured metallicity

\def\edcomment#1{\iffalse\marginpar{\raggedright\sl#1\/}\else\relax\fi}
\reversemarginpar

\begin{document}
\title{The Metallicity Evolution of Damped Lyman--$\alpha$ systems}
 \author{S. Savaglio \altaffilmark{1}, N. Panagia \altaffilmark{1}, 
M. Stiavelli \altaffilmark{1}}

\affil{\altaffilmark{1}
Space Telescope Science Institute, 3700 San Martin Drive,  Baltimore, MD
21218, USA; on assignment from the Space Science Department of the European
Space Agency}

\begin{abstract} We have collected  data for \nmea~Damped Lyman--$\alpha$ 
(DLA) systems, to investigate the chemical evolution of galaxies in
the redshift interval $0.0 < z < 4.4$.  In doing that, we have adopted
the most general approach used so far to correct for dust depletion.
The best solution, obtained through $\chi^2$ minimization, gives as
output parameters the global DLA metallicity and the dust--to--metals
ratio.  Clear evolution of the metallicity vs. redshift is found
(99.99\% significance level), with average values going from
$\sim1/30$ solar at $z\sim4.1$ to $\sim3/5$ solar at $z\sim0.5$.  We
also find that the majority of DLAs ($\sim60$\%) shows dust depletion
patterns which most closely resemble that of the warm halo clouds in
the Milky Way, and have dust--to--metals ratios very close to warm
halo clouds.

\end{abstract}

\section{Introduction}
 
Many models of galaxy formation and evolution in recent years
take as a reference for the observational side the results coming from
QSO absorption studies and in particular those objects that show a
large HI column density, namely, the damped Lyman--$\alpha$ systems
(DLAs) with $N_{HI} \gsim 10^{20}$ atoms \cm2 (see for example Pei et
al. 1999). DLAs have been widely studied both because they are
believed to be the progenitors of present day galaxies and because the
large HI column density allows one to probe the chemical enrichment
problem.  DLAs constitute so far the best laboratory where to directly
measure the heavy element enrichment in a large interval of
evolutionary phases, and to understand the processes of star formation
and metal pollution of the Universe. However, this kind of
investigation requires a careful consideration of the effects of dust
depletion suffered by DLAs (Pei et al. 1991; Pettini et al. 1994).  We
present the analysis of a sample of \nmea~ DLAs in the redshift range
$0.0 < z < 4.4$ to investigate their chemical state. We find that,
after allowance for dust depletion corrections which are obtained
with a very general approach, the DLA population clearly shows a
metallicity--redshift evolution.

\section{Chemical abundances and evolution}

\begin{table}[h]
\begin{scriptsize}
\caption{Mean element abundances relative to hydrogen}
\begin{tabular}{cccccccc}
\hline\hline&&&&&&&\\[-4pt] 
        &   \multicolumn{3}{c}{[X/H]} &&  \multicolumn{3}{c}{[X/Fe]} \\
[4pt]\cline{2-4}\cline{6-8}\\
Element & DLA & Warm Halo$^{a}$ &  SMC$^b$ && DLA & Warm Halo$^{a}$ &  SMC$^b$ \\
[5pt]\hline&&&&&\\[-5pt] 
Mn & $-1.26\pm0.51$ & $-0.60\pm0.13$ & $-1.10^{+0.14}_{-0.12}$ && $-0.08\pm0.21$ & $+0.04$   & $+0.11^{+0.04}_{-0.05}$ \\
Zn & $-0.96\pm0.44$ & --             & $-0.64^{+0.16}_{-0.14}$ && $+0.58\pm0.29$ & -- & $+0.57^{+0.08}_{-0.09}$ \\
Cr & $-1.36\pm0.45$ & $-0.51\pm0.13$ & $-1.13^{+0.14}_{-0.11}$ && $+0.17\pm0.09$ & $+0.13$   & $+0.08^{+0.03}_{-0.04}$\\
Fe & $-1.61\pm0.48$ & $-0.64\pm0.06$ & $-1.21^{+0.14}_{-0.11}$ && -- & -- & -- \\
Ni & $-1.66\pm0.45$ & $-0.84\pm0.07$ & $-1.73^{+0.17}_{-0.16}$ && $-0.14\pm0.19$ & $-0.21$   & $-0.52^{+0.09}_{-0.11}$  \\
Si & $-1.28\pm0.58$ & $-0.28\pm0.19$ & $-0.57^{+0.14}_{-0.11}$ && $+0.43\pm0.18$ & $+0.36$   & $+0.64^{+0.04}_{-0.04}$\\
S  & $-1.37\pm0.42$ & $-0.04\pm0.20$ & -- && -- & -- & --  \\
[2pt]\hline\\
\end{tabular}
$a)$ Savage \& Sembach (1996); $b)$ Measured toward Sk 108, Welty et al. 
(1997).
\end{scriptsize}
\end{table}

We have collected data from the literature for a DLA sample, which
includes \nmea~objects.  This sample represents the largest and most
complete sample of DLAs for which measurements of HI and
heavy--element column densities are available. The ions considered for
abundance measurements are FeII, ZnII, CrII, SiII, MnII, SII, NiII.
These ions are the dominant contributors to the abundances of the
corresponding elements in HI clouds with high column densities,
because they all have ionization potentials below 13.6~eV.

In Table~1 we give the mean metal abundances relative to hydrogen and
iron. They are presented with the customary definition [X/H]$_{DLA} =
\log ({\rm X}_i/{\rm Y}_i)_{DLA} - \log (\rm X/Y)_\odot$, where X$_i$
and Y$_i$ are the ion column densities of element X and Y. For
comparison, the mean abundances for Warm Halo (WH) clouds (Savage and
Sembach 1996) and the Small Magellanic Cloud (SMC, Welty et al. 1997)
are also shown.  We note that globally DLAs show [X/H] and [X/Fe]
abundance ratios more similar to those of SMC and WH clouds,
respectively.  This suggests that metal abundances in DLAs are the
result of chemical enrichment processes similar from the ones
operating in the SMC and that the most common depletion pattern
operating in DLAs is similar to the one observed in WH clouds.
 
Indeed, to derive a complete picture of the DLA chemical state, one
must correct for dust depletion effects.  Since every element
considered is affected by dust depletion differently, one must
consider all measured species simultaneously.  In the Milky Way, a
number of depletion patterns have been identified, showing highest
depletions in dense disk clouds and lowest depletions in low density,
warm halo clouds (Savage \& Sembach 1996).  We make a simplification
assuming that the depletion patterns in DLAs may be reproduced by one
of the four depletion patterns identified for the MW: Warm Halo, Warm
Halo + Disk (WHD), Warm Disk (WD) and Cool Disk (CD) clouds (Savage \&
Sembach 1996), thus modifying the dust--to--metals ratio to obtain the
best match with the observations.  By means of a $\chi^2$ minimization
procedure we determine the best fit DLA metallicities and the
dust--to--metals ratios.
 
\begin{figure}[h]
\plotfiddle{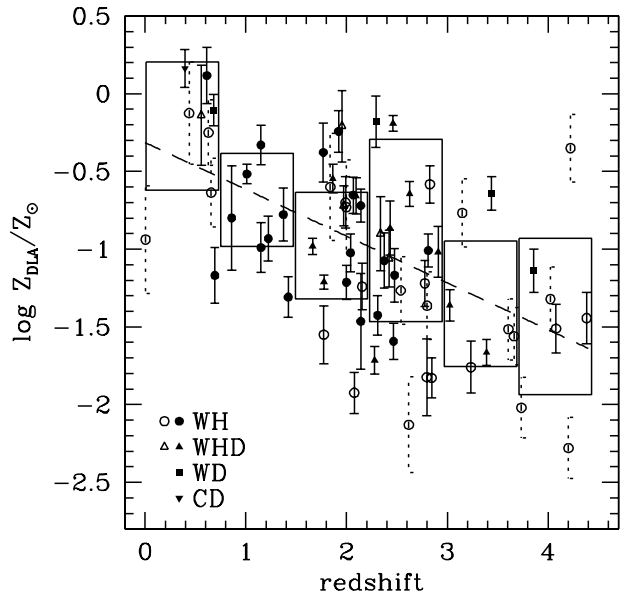}{8cm}{0}{85}{85}{-180}{-150}
%\plotfiddle{fig1.ps}{13cm}{0}{40}{40}{-120}{-60}
\caption{\label{f1} Total (gas+dust) metallicity compared to solar for
\nmea~ DLAs as a function of redshift.  Filled symbols represent the
37 DLAs with 3 or more measured elements, open symbols are DLAs with
two (solid bars, 16 objects) or one (dotted bars, 16 objects) element
measured. The dashed line is the best fit in the case of linear corelation
and that gives a slope $d\log
Z_{DLA}/dz=-0.30\pm 0.06$.}
%lines are the model predictions of global metal
%production of the Universe obtained from the star formation history
%(solid line, Madau \& Pozzetti, 1999), from hydrodynamical
%cosmological simulations (dotted line, Cen \& Ostriker, 1999) and from
%DLA evolution (dashed line, Pei \& Fall, 1995).  Boxes are centered on
%the weighted averages over $\Delta z=0.74$ intervals and vertical
%widths mark the corresponding $\pm1\sigma$ weighted dispersion.}
\end{figure} 

Fig.~1 shows the metallicity as a function of redshift. Filled symbols
represent DLAs with three or more measured elemental abundances for
which it has been possible to obtain a proper best fit solution (37
DLAs).  For the cases with only two elements observed, each fit has a
zero formal error and, therefore, a reduced $\chi^2$ cannot be
calculated; thus, the best fit is considered less significant (16
DLAs, empty symbols and solid error bars).  Finally, for the cases
where only one element is measured, we estimate the metallicity
assuming a WH depletion pattern (16 DLAs, empty symbols and dotted
error bars).  The combination of the largest sample available (\nmea~
DLAs), a large redshift baseline ($0.0<z<4.4$) and a more accurate
dust correction applied have led to the unambiguous detection of the
redshift evolution of metallicity in DLA galaxies, with mean values
around 1/30 of solar at $z\sim4.1$ to 3/5 of solar at $z\sim0.5$. We
found a significant linear correlation of metallicity with redshift
(99.99\% significance level) with a slope of $-0.301\pm0.056$, which
is clearly not consistent with a null gradient, indicating genuine
evolution of DLA metallicity with redshift.  In Fig.~1 we also show
six boxes centered on the weighted averages over $\Delta z=0.74$
intervals and whose vertical widths mark the corresponding
$\pm1\sigma$ weighted dispersion. In addition, we note that the
vertical dispersion of points greatly exceeds the total errors,
indicating that, although all DLAs align along an average trend, there
are real differences among individual objects in either initial
conditions, or time of formation, or metal enrichment efficiency, or
all of the above.

Pei \& Fall (1995) consider that a mean heavy--element abundance in
the interstellar medium of DLA galaxies is given by the ratio of the
total metal content to the total gas content ({\i.e.}
$Z\propto~\Sigma_j~N_{metals,j}/\Sigma_j N_{HI,j}$), which means that
large HI DLAs dominate when calculating the global metallicity.  This
kind of analysis has been performed on a sample of DLAs using the ZnII
absorption line and a null result has been found for the evolution
(Pettini et al. 1999), and it is not disproved if our sample is
used.  However, the lack of evident evolution in this case appears to
be due to the fact that those DLAs with large HI column density are
concentrated in the central redshift region (84\% of DLAs with $\log
N_{HI}>20.7$ are in the bin $1.4 < z < 3.5$). In other words, we
cannot exclude the redshift evolution of metallicity in high HI DLAs
till when the considered sample will be fairly distributed in the
considered redshift range. If there were a metallicity evolution also
for high HI DLAs, dust obscuration could be too large at $z<1$ to make
DLAs detectable (a DLA with $\log N_{HI}=21.0$ and metallicity $\log
Z_{DLA}/Z_\odot=-0.5$ would extinct about 1.5 magnitude in the
Ly$\alpha$). Moreover, this is where UV observations, required to
detect Ly$\alpha$ absorption, are very sensitive to any appreciable
dust opacity.

In 37 DLAs for which we have obtained a best fit solution, we have
found that the majority ($\sim 60$\%) show a WH--like depletion
pattern, then $\sim 30$\% and $\sim 10$\% have a WHD--like and
WD--like depletion pattern, respectively.  The predominance of
WH--like conditions may be interpreted as evidence for a lower density
in the absorbing clouds relative to clouds in the MW disk and/or to
more efficient dust destruction processes operating in DLA galaxies,
possibly due to more active star formation that may produce a more
turbulent, dust hostile environment.  In our sample there are 31 DLAs
with measured Si and for which it has to perform $\chi^2$
minimization.  For 8 over 31 systems, there is evidence of deviation
of Si with respect the expectations, having a mean value of
$0.28\pm0.05$. No correlation with redshift and metallicity is
revealed. This result may be indicative of $\alpha$--element
enrichment in a fraction of DLAs, or of a lower efficiency of silicate
formation than is found in the Milky Way.

We like to stress that it is the combination of a large sample of DLAs
(\nmea~objects), a wide redshift baseline ($0.0<z<4.4$) and a more
accurate dust correction applied to all heavy--elements measured, that
has allowed us to demonstrate that the metallicity of DLAs evolves
with redshift. The combined use of as many elements as possible allows
one to much more efficiently track the metallicity evolution down to
very metal poor gas clouds.

\end{document}